\documentclass[aps,prl,showpacs,floatfix,twocolumn,amsmath,amssymb,superscriptaddress]{revtex4-1}
\usepackage{graphicx}
\usepackage{color}

\newcommand{\Ang}{\AA$^{-1}$}

\newcommand{\malah}{Cu$_2$(OH)$_2$CO$_3$}
\newcommand{\malad}{Cu$_2$(OD)$_2$CO$_3$}
\newcommand{\q}{\ensuremath{ {\bf Q} }}

\begin{document}
\title{Spin excitations in the two-dimensional strongly coupled dimer system malachite}
\author{E. Can\'evet}
   \affiliation{Laboratoire de Physique des Solides, Universit\'e Paris Sud 11, UMR CNRS 8502, F-91405 Orsay, France}
   \affiliation{Institut Laue-Langevin, CS 20156, F-38042 Grenoble Cedex 9, France}
\author{B. F\aa k}
   \affiliation{Institut Laue-Langevin, CS 20156, F-38042 Grenoble Cedex 9, France}
   \affiliation{Univ.\ Grenoble Alpes, INAC-SPSMS, F-38000 Grenoble, France}
   \affiliation{CEA, INAC-SPSMS, F-38000 Grenoble, France} 
\author{R. K. Kremer}
   \affiliation{Max-Planck Institute for Solid State Research, Heisenbergstrasse 1, 70569 Stuttgart, Germany}
\author{J.~H.~Chun}
  \affiliation{Max-Planck Institute for Solid State Research, Heisenbergstrasse 1, 70569 Stuttgart, Germany}
\author{M.~Enderle}
   \affiliation{Institut Laue-Langevin, CS 20156, F-38042 Grenoble Cedex 9, France}
\author{E.~E.~Gordon}
   \affiliation{Department of Chemistry, North Carolina State University, Raleigh, North Carolina 27695-8204, USA}
\author{J. L. Bettis}
   \affiliation{Department of Chemistry, North Carolina State University, Raleigh, North Carolina 27695-8204, USA}
\author{M.-H.~Whangbo}
   \affiliation{Department of Chemistry, North Carolina State University, Raleigh, North Carolina 27695-8204, USA}
\author{J. W. Taylor}
   \affiliation{ISIS Facility, Rutherford Appleton Laboratory, Chilton, Didcot, Oxon OX11 0QX, United Kingdom}
\author{D. T. Adroja}
   \affiliation{ISIS Facility, Rutherford Appleton Laboratory, Chilton, Didcot, Oxon OX11 0QX, United Kingdom}

\date{\today}

\begin{abstract}
The mineral malachite, \malad, 
has a quantum spin-liquid ground state 
and no long-range magnetic order down to at least $T=0.4$~K. 
Inelastic neutron scattering measurements show that the excitation spectrum consists of dispersive gapped singlet-triplet excitations, 
characteristic of spin-1/2  dimer-forming Heisenberg antiferromagnets.
We identify a new two-dimensional dimerized coupling scheme with strong interdimer coupling $J^\prime/J_1\approx 0.3$ 
that places malachite between strongly coupled alternating chains, square lattice antiferromagnets, and infinite-legged ladders. 
The geometry of the interaction scheme resembles the staggered dimer lattice, which may allow unconventional quantum criticality.    
\end{abstract}

\maketitle

Entangled quantum systems are exciting in view of their potential impact on quantum computing. 
A macroscopic number of entangled qubits is embodied in two-dimensional (2D) spin-1/2 antiferromagnets with spin-singlet quantum ground states. 
These systems are equally fascinating on a fundamental level, as they display a variety of generic many-body quantum phenomena and new types of excitations.
Well-known examples are  
the Shastry-Sutherland model \cite{Shastry81}, 
the Kitaev-honeycomb model with flux-type and Majorana-fermion excitations \cite{Kitaev06,Knolle14}, 
and the spin-liquid ground state of the frustrated square lattice antiferromagnet \cite{Dagotto89}. 

Referring to a unit of two entangled $S=1/2$ spins as a dimer, the simplest 2D case consists of weakly coupled dimers, 
where the excitations may be understood as propagating triplets resulting from a broken dimer bond. 
Hard-core boson models capture most of the related phenomena \cite{Mikeska06,James08,Tennant12,Quintero12,Ruegg05,Giamarchi08}.  
The intermediate and strong coupling cases, however, are less well understood. 
Deconfined fractional quasi-particles may emerge at the quantum critical point to antiferromagnetic order \cite{Senthil04},
as predicted for the frustrated square lattice \cite{Jiang12}. 
Near the quantum critical point, 
partial fractionalization or the formation of more extended entangled states 
may play a role, as predicted for the zero-field excitation spectrum  
and the magnetized states of the Shastry-Sutherland model \cite{Totsuka01,Corboz14}. 

\begin{figure}[b]
\includegraphics[width=.99\columnwidth]{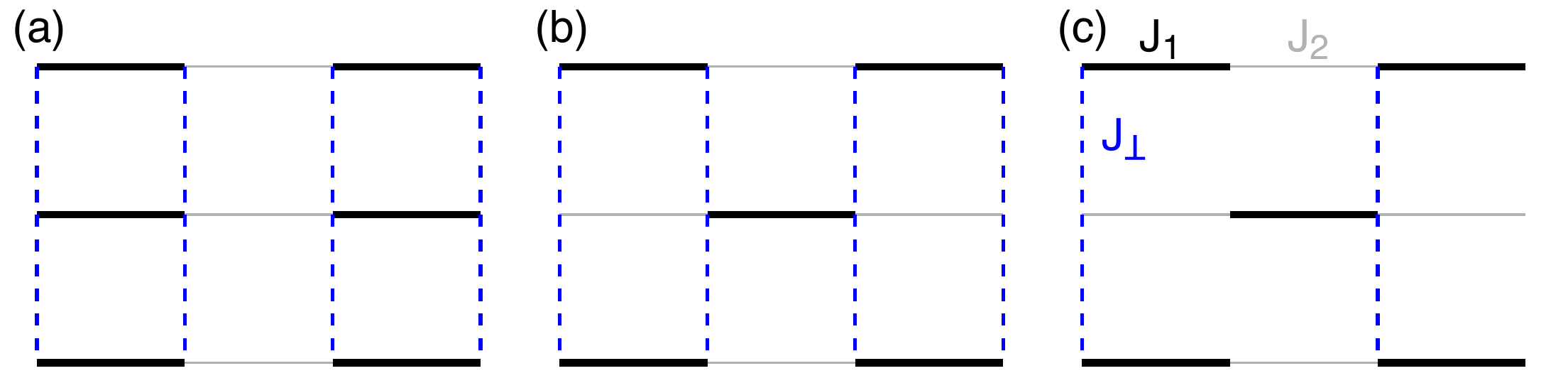}
\caption{(Color online) 
Schematic view of dimer lattices. (a) Columnar, (b) staggered, and (c) depleted staggered.  
Thick black lines show the dimer bonds $J_1$, 
thin grey lines the intrachain interactions $J_2$, 
and (blue) dashed lines the interchain interactions $J_\perp$. 
}
\label{FigDimerLattice}
\end{figure}

Commonly, quantum many-body phenomena in 2D are associated with the presence of frustrated interactions. 
However, it was recently shown that certain classes of non-frustrated 2D coupled dimers develop non-trivial  cubic interactions of purely quantum mechanical origin 
that lead to simultaneous condensation of the one-triplon and two-triplon excitations at the $\Gamma$-point at the critical interdimer coupling strength \cite{Fritz11}. 
These three-particle interactions have been shown to correspond to Berry-phase terms in the O(3) non-linear $\sigma$-model \cite{Fritz11} 
and lead to the equivalent of vector boson excitations in the adjacent N\'{e}el phase \cite{Huh13}. 
The geometry of the dimer lattice has been shown to be crucial: While the columnar dimer lattice (Fig.\ \ref{FigDimerLattice}a) displays a conventional quantum critical point (QCP), 
the staggered dimer lattice (Fig.\ \ref{FigDimerLattice}b) has a QCP with dangerously irrelevant Berry-terms  \cite{Fritz11}. 

Experimental realizations of systems that could confirm these theoretical ideas are scarce. 
In this Article, we have studied the magnetic excitations in the mineral malachite, 
which realizes a variant of the staggered dimer lattice, 
where half of the interdimer couplings perpendicular to the dimers are removed (see Fig.\ \ref{FigDimerLattice}c). 
Similar to the staggered dimer lattice,  
the dimers in this depleted staggered dimer lattice are coupled in an asymmetric fashion,  
which leads to a QCP with emergent antiferromagnetism at the $\Gamma$-point ${\bf k}=0$.
The malachite coupling scheme therefore falls in the category of models where Berry-phase terms may be important at the QCP \cite{Fritz11}.

\begin{figure}
\includegraphics[width=.8\columnwidth]{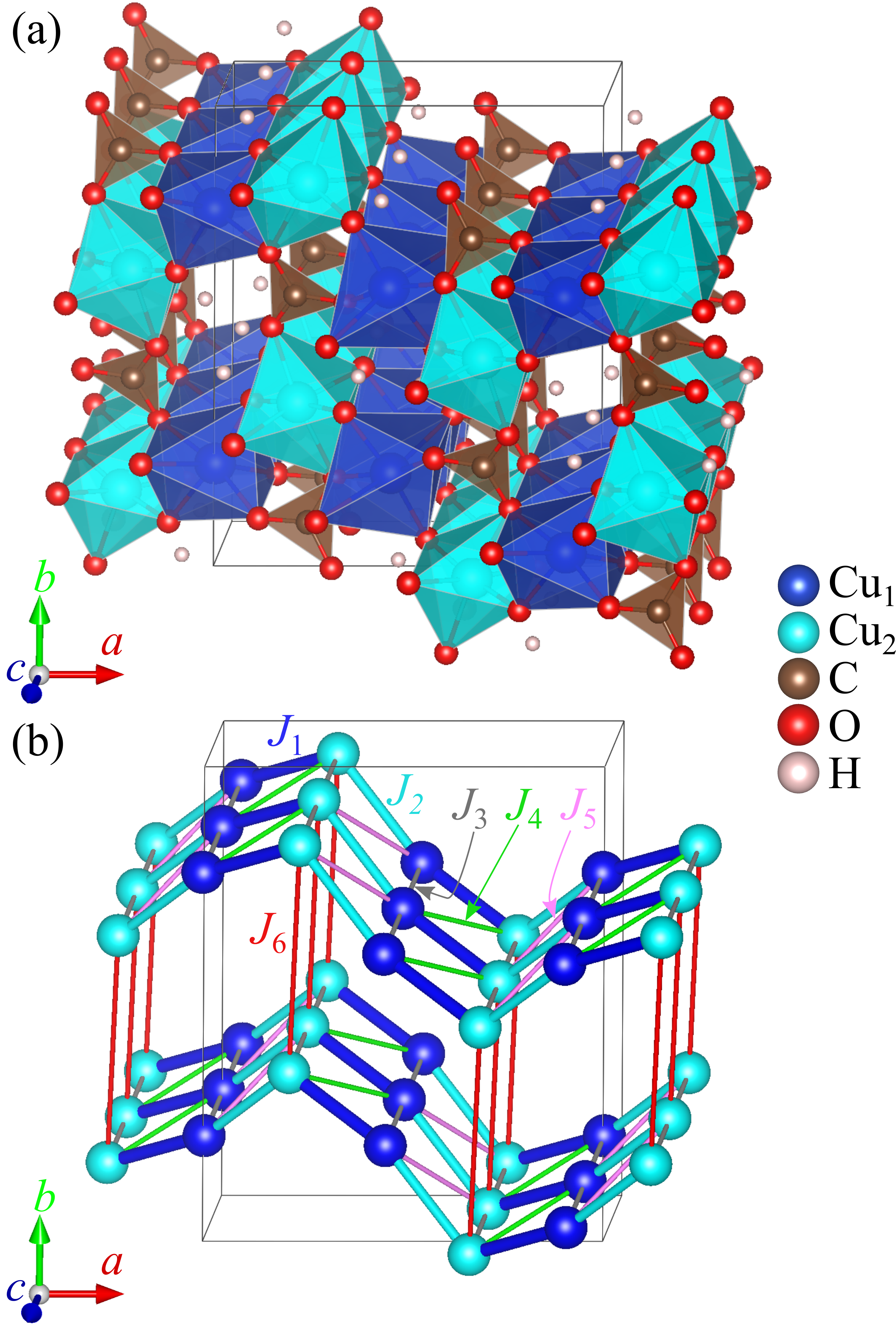}
\caption{(Color online) (a)~Crystal structure of malachite, \malah. 
(b)~Schematic structure showing the buckled dimer chains of the spin-1/2 Cu atoms 
and the exchange interactions $J_1$--$J_6$. }
\label{FigCrystal}
\end{figure}

Malachite is a natural mineral with chemical formula \malah\ that has been used as a green pigment and gem stone since the antiquity, 
but whose magnetic properties only recently have attracted attention. 
It crystallizes in the monoclinic space group $P2_1/a$ (No.\ 14) with lattice parameters $a=9.50$, $b=11.97$, $c=3.24$~\AA, 
and monoclinic angle $\beta=98.8^\circ$ \cite{Zigan77,Lebernegg13}. 
The crystal structure is shown in Fig.\ \ref{FigCrystal}a. 
The Cu$^{2+}$ ions form dimerized spin-1/2 chains running in the $a-c$ plane, see Fig.\ \ref{FigCrystal}b.  
The chains are built up by an alternation of two identical but differently oriented dimers, 
leading to a buckling of the chains. 
The unit cell therefore contains four dimers rather than two for the unbuckled case.

\begin{figure}
\includegraphics[width=.98\columnwidth]{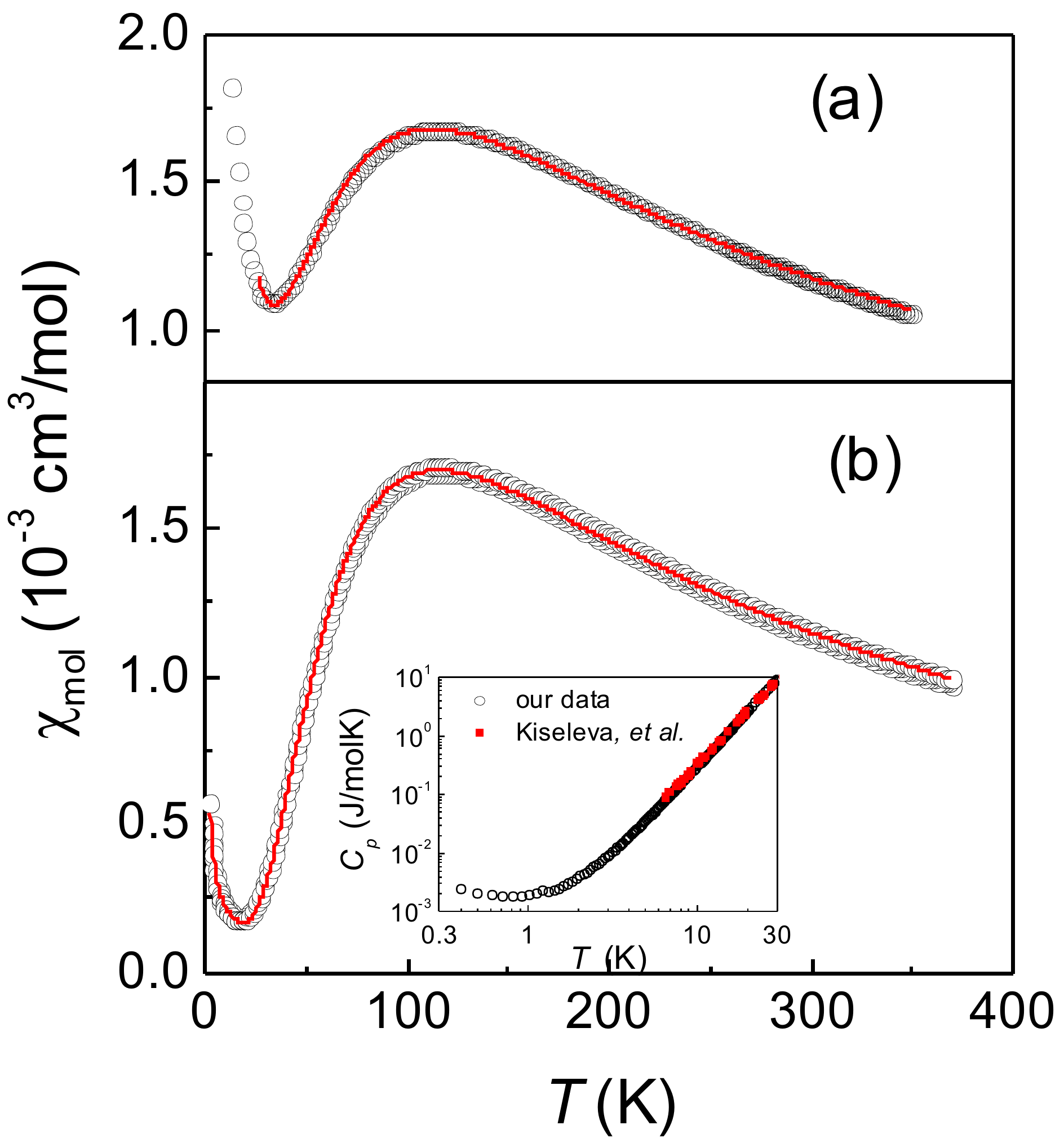} 
\caption{(Color online)
Magnetic susceptibility of synthesized powder of \malad\  (a) 
and of  a natural crystal of \malah\ (b).  
The data of the former is multiplied by a factor of 1.13 to compensate for diamagnetic impurity phases (see Ref.\ \cite{Suppl}). 
The (red) line is a fit to a high-temperature series expansion of an alternating chain model (see text). 
Inset: Specific heat of the natural \malah\ crystal showing the absence of magnetic order down to $T=0.4$~K. 
Open (black) circles are our data and filled (red) squares are data at higher temperatures from Ref.\ \cite{Kiseleva92}. }
\label{FigX}
\end{figure}

Magnetic  susceptibility measurements on natural mineral samples of \malah\ show absence of magnetic order down to $T=2$~K 
and a spin gap of 130~K \cite{Lebernegg13,Janod00}. 
The susceptibility data is well described by an alternating $S=1/2$ antiferromagnetic chain model \cite{Johnston00}  
with an intradimer exchange of $J_1= 190$~K and an interdimer exchange $J_2=90$~K  \cite{Lebernegg13,Janod00}. 
This is confirmed by our magnetic  susceptibility measurements on a high-quality natural crystal of \malah\ 
with a magnetic field of 1~T along the $c$-axis (see Fig.\ \ref{FigX}b and Supplemental Material \ \cite{Suppl}). 
Specific-heat measurements on this crystal \cite{Suppl}
show the absence of long-range magnetic order 
down to temperatures as low as $T=0.4$~K (see inset of Fig.\ \ref{FigX}b). 
The specific heat at higher temperatures is in good  agreement with previous measurements  \cite{Kiseleva92}. 
The absence of spin freezing down to $T=2$~K  is also evidenced by ac-susceptibility measurements  \cite{Janod00}. 

\begin{figure}
\includegraphics[width=.99\columnwidth]{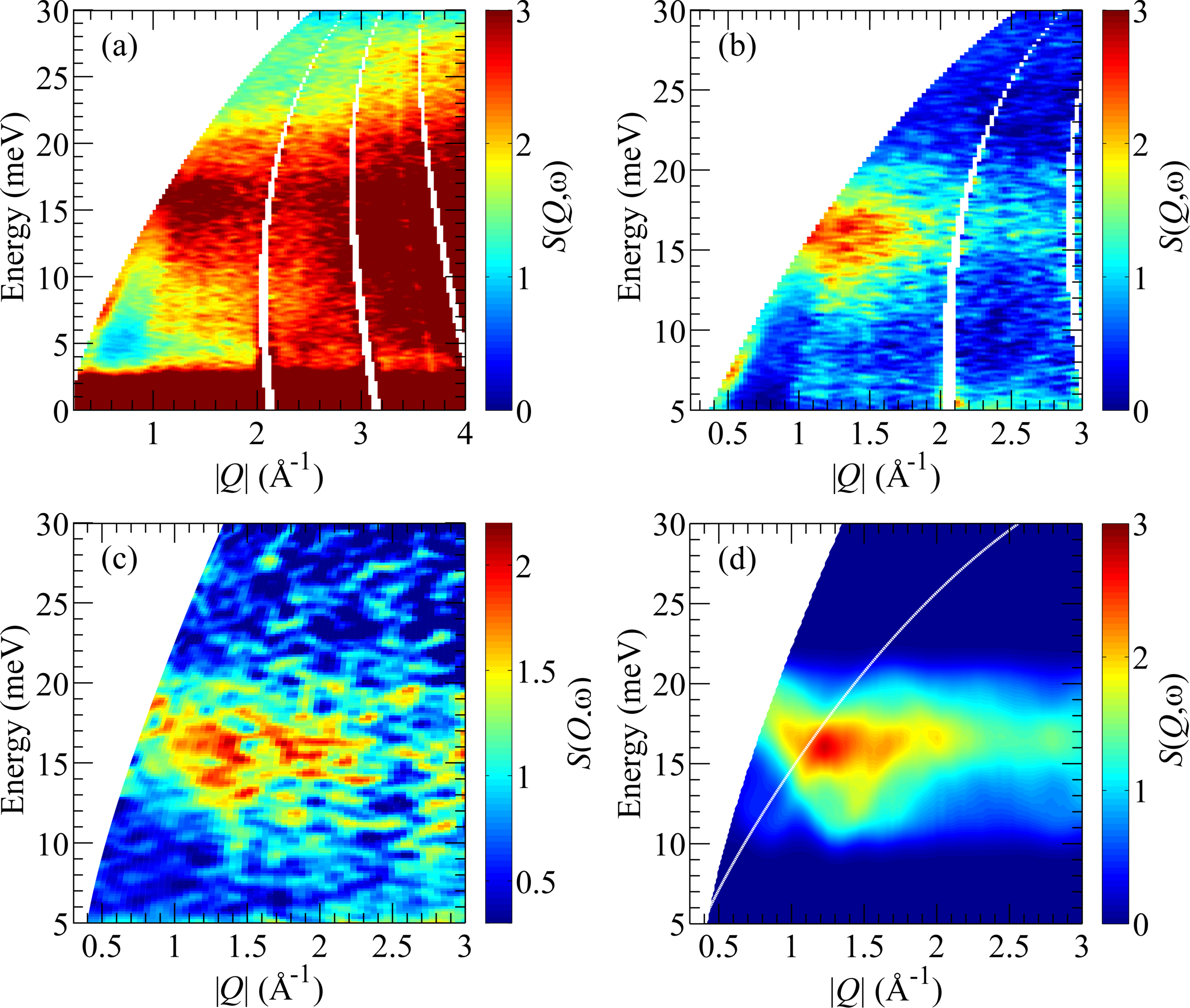}
\caption{(Color online) (a) Neutron scattering intensity $S(Q,E)$ on a logarithmic intensity scale as a function of wave vector $Q$ and energy $E $ 
of fully deuterated malachite \malad\  at a temperature of 5.9~K measured with an incoming neutron energy of $E_i=35$~meV. 
The magnetic scattering is centered at energies of 10--20 meV for $Q<2$~\Ang. 
(b) Magnetic part of $S(Q,E)$ after subtraction of the phonon contribution, measured at $E_i=35$~meV. 
(c) Same as (b) for $E_i=80$~meV.
(d)~Powder average of model calculations with $J_1=16.5$, $J_2=5.6$, and $J_6=4.3$~meV (see text). 
The white line indicates the accessible $Q$--$E$ range for $E_i=35$~meV. }
\label{FigMap}
\end{figure}

In order to obtain high-quality inelastic neutron scattering data, 
we prepared a fully deuterated powder sample of  \malad\ \cite{Suppl}. 
X-ray powder diffraction confirmed that the deuterated \malad\ sample has the same crystallographic structure as natural \malah\ samples  \cite{Suppl}.  
The degree of deuteration was found to be in excess of 95\% from Raman scattering of the OH-stretching mode \cite{Suppl}. 
The magnetic susceptibility of the deuterated sample is fully consistent with that of the natural hydrogenated samples  (see Fig.~\ref{FigX}).

Density functional theory (DFT+U) calculations by Lebernegg {\it et al.}\ \cite{Lebernegg13} 
and by us \cite{Suppl} 
reproduce the intra- and interdimer exchange couplings $J_1$ and $J_2$ inferred from magnetic susceptibility measurements \cite{Lebernegg13,Janod00}. 
The strongest coupling between dimer chains is $J_\perp\equiv J_6$ along the $b$ axis
(see Fig.\ \ref{FigCrystal}b for definition of the exchange integrals), 
which occurs via the carbonate CO$_3$ groups. 
The calculated $J_6$ is antiferromagnetic and of similar magnitude to $J_2$.

Inelastic neutron scattering measurements were performed for temperatures between 5.8 and 300~K on the MARI direct-geometry time-of-flight chopper spectrometer at ISIS  
using incident energies of $E_i = 20$, 35, 80, and 200~meV.
The corresponding energy resolution at elastic energy transfer was 0.53, 1.16, 2.37, and 7.84~meV.  
The fully deuterated powder sample of mass 16.2~g was wrapped in a thin aluminium foil in an annular geometry and thermalized by helium exchange gas. 
Standard data reduction procedures were used to transform the time-of-flight data to the dynamic structure factor $S(Q,E)$ \cite{mantid}, 
which is a function of wave vector $Q$ and energy transfer $E$. 
Measurements at high incoming energies (200~meV) show molecular modes of deuterium but not of hydrogen. 
This confirms the high deuteration rate inferred from Raman data. 
Low-temperature neutron scattering data taken at lower incoming energies show a band of magnetic scattering at energies 10--20 meV and low wave vectors $Q$.
This scattering disappears with increasing temperature and increasing wave vector,  confirming its magnetic origin. 
The total (magnetic plus nuclear) scattering as a function of $Q$ and $E$ measured with $E_i=35$~meV is shown in Fig.\ \ref{FigMap}a. 
At large $Q$ values, $S(Q,E)$ is dominated by phonon scattering. 
Since the phonon scattering is essentially incoherent with a smooth $Q$ dependence, 
it can be subtracted using established methods \cite{Fak08}. 
The resulting magnetic scattering is shown in Fig.\ \ref{FigMap}b. 
Efficient phonon subtraction can also be obtained by subtracting high-temperature data from low-temperature data. 
The resulting magnetic scattering is shown in Fig.\ \ref{FigMap}c for a higher incoming energy, $E_i=80$~meV, which allows access to lower $Q$ values.

\begin{figure}
\includegraphics[width=.99\columnwidth]{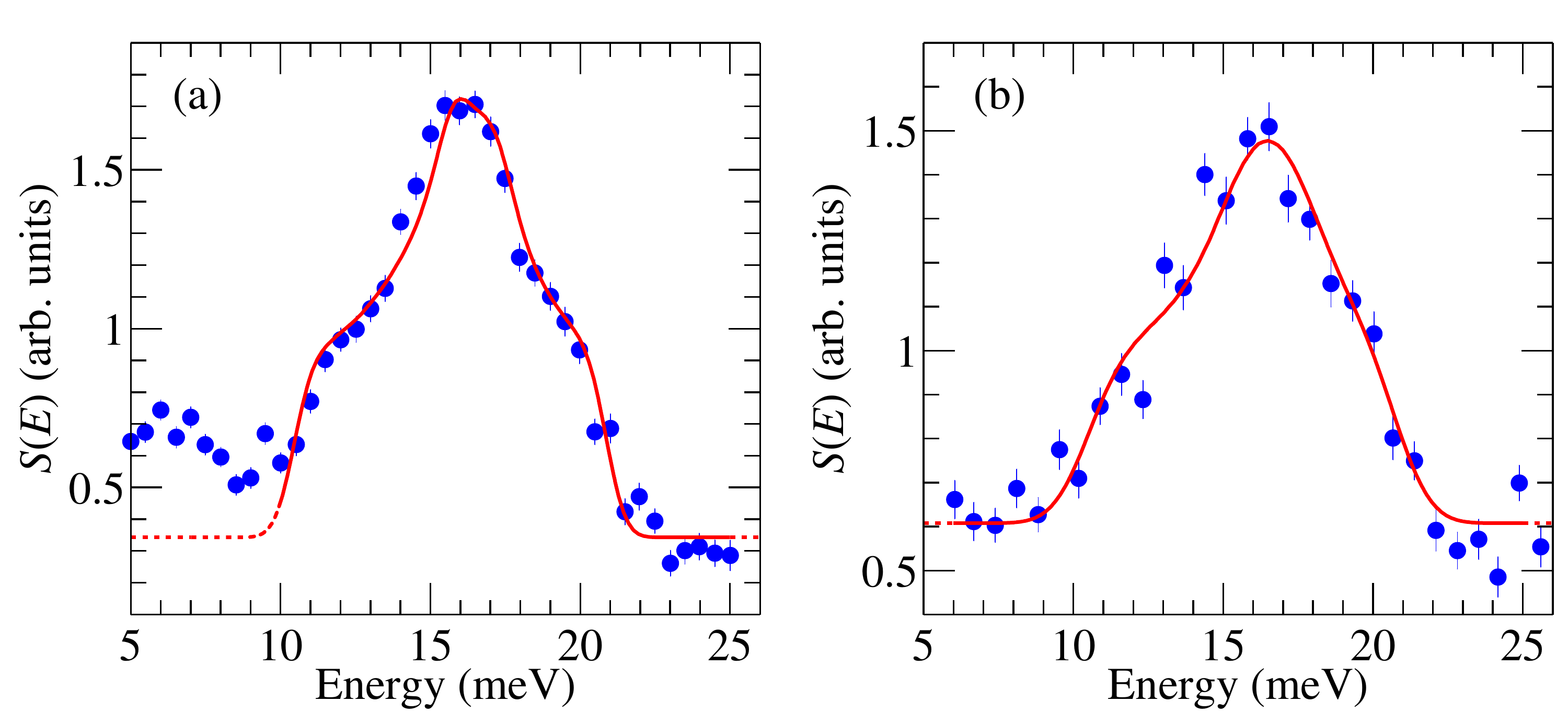}
\caption{(Color online) 
Energy dependence of the magnetic scattering integrated over the $Q$ range $0.5<Q<2.5$~\Ang\ for incoming energies of (a) $E_i=35$~meV and (b) $E_i=80$~meV.  
The lines show model calculations with $J_1=16.5$, $J_2=5.6$, and $J_6=4.3$~meV.  
}
\label{FigEscan}
\end{figure}

Qualitative information can be extracted directly from the bandwidth of the magnetic scattering shown in Fig.\ \ref{FigMap}b or \ref{FigMap}c. 
The lower and upper boundaries can be estimated to $E_-=11$ and $E_+=20$~meV, respectively. 
For a pure alternating chain without interchain interactions, 
this gives exchange interactions $J_1=(E_++E_-)/2=15.5$~meV and $J_2=E_+-E_-=9$~meV, and hence $J_2/J_1=0.58$. 
This is in reasonable agreement with the estimates 
of 16.4 and 7.8~meV
from the magnetic susceptibility fits discussed above. 
Using a first-moment analysis along the lines of Ref.~\cite{Stone02}, 
we find that $J_6$ is the dominating interchain interaction, as also indicated by DFT calculations \cite{Lebernegg13,Suppl}.

Magnetic excitations in two-dimensional networks of weakly coupled dimers have successfully been studied using linked cluster expansions \cite{Muller00,Muller03}. 
However, since the unit cell of malachite contains four dimers, such an approach is not straightforward. 
We therefore use the harmonic triplon approximation \cite{Fritz11} for a quantitative analysis of the magnetic excitations in malachite. 
This method is based on the bond-operator representation of the singlet-triplet excitations \cite{Sachdev90}. 
Using the formalism developed in Ref. \cite{Tsallis78}, this leads in the present case  to a non-Hermitian $8 \times 8$ matrix. 
From this matrix, the triplon energies $E_s({\bf Q})$ and spectral weights $I_s({\bf Q})$ are readily obtained by numerical methods. 
For a single crystalline sample, the dynamic structure function is given by 
\begin{equation}
S(\q,E) = 
 \frac{1}{1-e^{-E/k_BT}} 
\sum_s I_s(\q)\delta[E-E_s(\q)] , 
\label{EqSingle}
\end{equation}
where the index $s$ is over the 8 magnetic modes, of which four have positive energies. 
The powder-averaged dynamic structure function is given by 
\begin{eqnarray}
S(Q,E) &=& \frac{1}{4\pi} \int_{{\bf Q}=|Q|} \! {\rm d}\Omega \, S({\bf Q},E) =\nonumber\\
&=& \frac{1}{4\pi} \int_0^{2\pi}\! {\rm d}\phi \int_0^\pi {\rm d}\theta \sin\theta \, S({\bf Q}\!=\!|{\bf Q}|,E) , 
\label{EqSpow}
\end{eqnarray}
where the spherical average is readily evaluated using standard Monte-Carlo techniques. 
The calculated spectra for various combinations of exchange integrals are then compared with the data of Fig.\ \ref{FigMap}b or c. 
The calculated $S(Q,E)$ is multiplied by the square of the magnetic form factor to match the observed neutron scattering intensity. 
To determine the exchange constants, 
a very sensitive method is to compare a density-of-states-like $S(E)$ obtained by integrating $S(Q,E)$ over a suitable $Q$ range, as shown in Fig.\ \ref{FigEscan}, 
and calculate the $\chi^2$ values for various sets of $J_1$, $J_2$, and $J_6$. 
This method allows us to conclude that $J_6$ is the dominating interchain interaction 
and of similar strength to the intrachain interaction $J_2$.  
The best agreement with the experimental data is obtained for an intradimer coupling of $J_1=16.5$~meV and interdimer couplings of $J_2/J_1=0.34$ and $J_6/J_1=0.26$. 
The corresponding $S(Q,E)$ is shown in Fig.\ \ref{FigMap}d and the $Q$-integrated energy distributions in Fig.\ \ref{FigEscan}. 

Our one-triplon analysis of the magnetic excitation spectrum in malachite shows that the interaction between dimers is dominantly a 2D coupling, 
a scenario which is also supported by DFT calculations. 
This 2D coupling scheme contains only antiferromagnetic exchange interactions, 
and closely resembles the staggered dimer model, 
since parallel (in-chain) and perpendicular interdimer couplings are nearly equal. 
Despite the removal of every second perpendicular interdimer bond, 
malachite remains in the class of asymmetrically coupled dimer models, 
where non-trivial cubic interaction terms in connection with emergent ${\bf k}=0$ antiferromagnetism imply that the Berry phase is dangerously irrelevant and may lead to a deconfined QCP \cite{Senthil04}. 

The critical coupling strength can be estimated by calculating for which exchange
the minimum energy of the triplon dispersion goes to zero. 
Within the harmonic triplon approximation, the columnar (Fig.\ \ref{FigDimerLattice}a) and staggered (Fig.\ \ref{FigDimerLattice}b) dimer models both have $J^{\prime}/J_1=1/3$ \cite{Fritz11}. 
This is to be compared to Quantum Monte Carlo calculations, where $J^{\prime}/J_1=0.524$ \cite{Wenzel09} and $J^{\prime}/J_1=0.397$ \cite{Wenzel08} for these two dimer models, respectively.  
Hence, it appears that the harmonic triplon approximation works surprisingly well even close to the critical point. 
For the depleted staggered dimer model (see Fig.\ \ref{FigDimerLattice}c), 
we obtain for the ``equal-coupling'' case $J_2=J_6=J^{\prime}$ a critical coupling strength of $J^{\prime}/J_1=1/2$ within the harmonic triplon approximation, 
which is considerably higher than for the other dimer models. 
For malachite, where the dimer chains are buckled (see Fig.\ \ref{FigCrystal}b), 
the critical coupling strength is $\tilde{J}_c=(J_2+J_6)/J_1=1$. 
Our analysis of the measured excitation spectrum in malachite gives $(J_2+J_6)/J_1=0.6$, 
which compares favorably with the DFT+U estimate of 0.8, 
and which explains the relatively large size of the observed energy gap and the absence of long-range magnetic order.

In conclusion, our inelastic neutron scattering measurements on a powder sample of the mineral malachite \malad\ demonstrate 
that it is a rare realization of a two-dimensional unfrustrated staggered dimer system with strong and near-equal parallel and perpendicular interdimer couplings. 
This allows for cubic triplon interactions in combination with emergent ${\bf k}=0$ antiferromagnetic order,  
which are essential characteristics required for Berry-phase terms leading to deconfined quantum critical points. 
Experimental studies of the spin dynamics of single crystalline malachite under pressure could therefore open the possibilities 
to investigate unusual features related to the presence of cubic triplon interactions, 
new types of quasiparticles emerging from partial fractionalization of the triplons, multi-triplon dynamics, 
and vector-boson excitations in the N\'eel phase near a deconfined QCP.

This work was supported in part by the ANR Grant No. ANR-12-BS04-0021. 
Experiment at the ISIS neutron and muon source was supported by a beam-time allocation from the Science and Technology Facilities Council, UK. 
We thank H.-J. Mikeska and T. Ziman for  helpful discussions 
and E. Br\"ucher, A. Schulz, and C. Stefani for experimental assistance.

\end{document}